\documentclass[aps,prl,reprint,floatfix]{revtex4-2}

\usepackage{amsmath}
\usepackage{amssymb}
\usepackage{mathtools}
\usepackage[bb=dsserif]{mathalpha}
\usepackage{bm}
\usepackage[per-mode=symbol, uncertainty-mode=compact-marker]{siunitx}
\DeclareSIUnit\bar{bar}
\DeclareSIUnit\amu{amu}
\DeclareSIUnit\eVpercSquared{\text{eV}/\text{\ensuremath{c^2}}}
\usepackage{isotope}
\usepackage{graphicx}
\usepackage[nomain,acronym,postdot,nogroupskip]{glossaries-extra}
\usepackage{longtable}
\usepackage{multirow}
\usepackage{booktabs}
\usepackage{xcolor}
\usepackage[hidelinks]{hyperref}

\begin{document}

\DeclareSIUnit\bar{bar}
\DeclareSIUnit\cps{cps}
\title{Magneto-optical trapping of zinc} 

\author{Lukas Möller}
\affiliation{Physikalisches Institut, Rheinische Friedrich-Wilhelms-Universität Bonn, Germany}
\author{Simon Stellmer}
\email{stellmer@uni-bonn.de}
\affiliation{Physikalisches Institut, Rheinische Friedrich-Wilhelms-Universität Bonn, Germany}
\date{\today}

\begin{abstract}
We report on laser cooling and magneto-optical trapping of atomic zinc. The atoms are cooled using the 213.9\,nm $^1$S$_0$ $\rightarrow$ $^1$P$_1$ transition, making this the shortest wavelength employed for magneto-optical trapping thus far. We demonstrate trapping of all stable isotopes of zinc, including the fermionic isotope $^{67}$Zn, which features a very narrow $^1$S$_0$ $\rightarrow$ $^3$P$_0$ transition that could form the basis of an optical atomic clock. We characterize the influence of various parameters on the MOT population and loading rate. The results presented here constitute the first step towards the application of zinc for high-precision optical spectroscopy and quantum information processing.
\end{abstract}

\maketitle

The magneto-optical trap (MOT) is one of the most widely used methods in modern atomic physics and usually provides the first cooling stage for optical atomic clocks \cite{Ludlow, SrClock}, atom interferometers \cite{Poli, Abe, Tino, Ruschewitz}, quantum gas experiments \cite{Kraft, Stellmer, Escobar, Chen}, and various efforts in the field of quantum computation and simulation \cite{Daley, Madjarov, Bloch, Schäfer, Gorshkov, Daley2, Bluvstein}. Here, elements with two valence electrons, generally referred to as alkaline earth(-like) metals, enjoy particular attention due to their long-lived metastable states and correspondingly sharp clock transitions. 

A subgroup of these species are the so-called group-IIB-elements (Zn, Cd, Hg, and Cn). The principal transitions of these elements are situated in the deep ultra-violet (DUV) region of the spectrum. Laser cooling and manipulation of these elements requires the development of powerful, tunable continuous wave (cw), DUV laser systems, which has already been achieved for Cd and Hg \cite{Kaneda, yamaguchi, Ohmae, Bandarupally, Gibble, McFerran, Lavigne, Zhang, ZhangX}. In return, these elements offer very broad dipole transitions for powerful cooling and fast read-out, as well as narrow intercombination lines for second-stage cooling. Their clock transitions are less sensitive to black-body radiation compared to elements typically used for optical lattice clocks, such as Sr and Yb \cite{Ludlow, Dzuba}. The short wavelengths allow for high spatial resolution in addressing and imaging dense arrays of individual atoms, a technique advantageous in neutral-atom quantum computing. For a long time the shortest wavelength used for magneto-optical trapping was the $229$\,nm $^1$S$_0$ $\rightarrow$ $^1$P$_1$ transition in Cd \cite{CdMOT}. This record was recently broken by the magneto-optical trapping of AlF using a $227.5$\,nm transition\cite{AlF}.

Zinc, with five stable bosonic isotopes (nuclear spins $I=0$) and one fermionic isotope ($^{67}$Zn, $I=5/2$), favorably combines many of the properties demanded of elements used in optical clocks, but has received little attention thus far due to the challenges in DUV laser development \cite{Wang}.

In this Letter, we present the first magneto-optical trap of atomic zinc. We demonstrate trapping of all five stable isotopes, perform a careful optimization of parameters, and show that ionization losses are not detrimental to laser cooling. Our work represents the first step in the exploration of zinc as a candidate for an optical clock or for quantum information processors. The cooling transition at $213.9$\,nm makes this the shortest wavelength used for magneto-optical trapping thus far.

\begin{figure}[b]
	\centering
	\includegraphics[width=\linewidth]{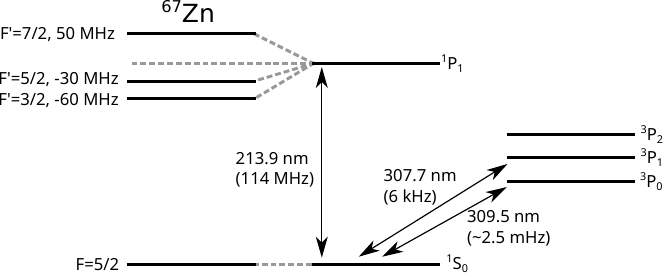}
	\caption{Partial level diagram of zinc, showing transitions relevant for laser cooling and precision spectroscopy, as well as their wavelengths $\lambda$ and natural linewidths $\Gamma/2\pi$. For the two states relevant to this work, the hyperfine splitting of the fermionic isotope $^{67}$Zn is shown as wll.}
	\label{fig:transitions}
\end{figure}

\begin{figure}[htpb]
	\centering
	\includegraphics[width=\linewidth]{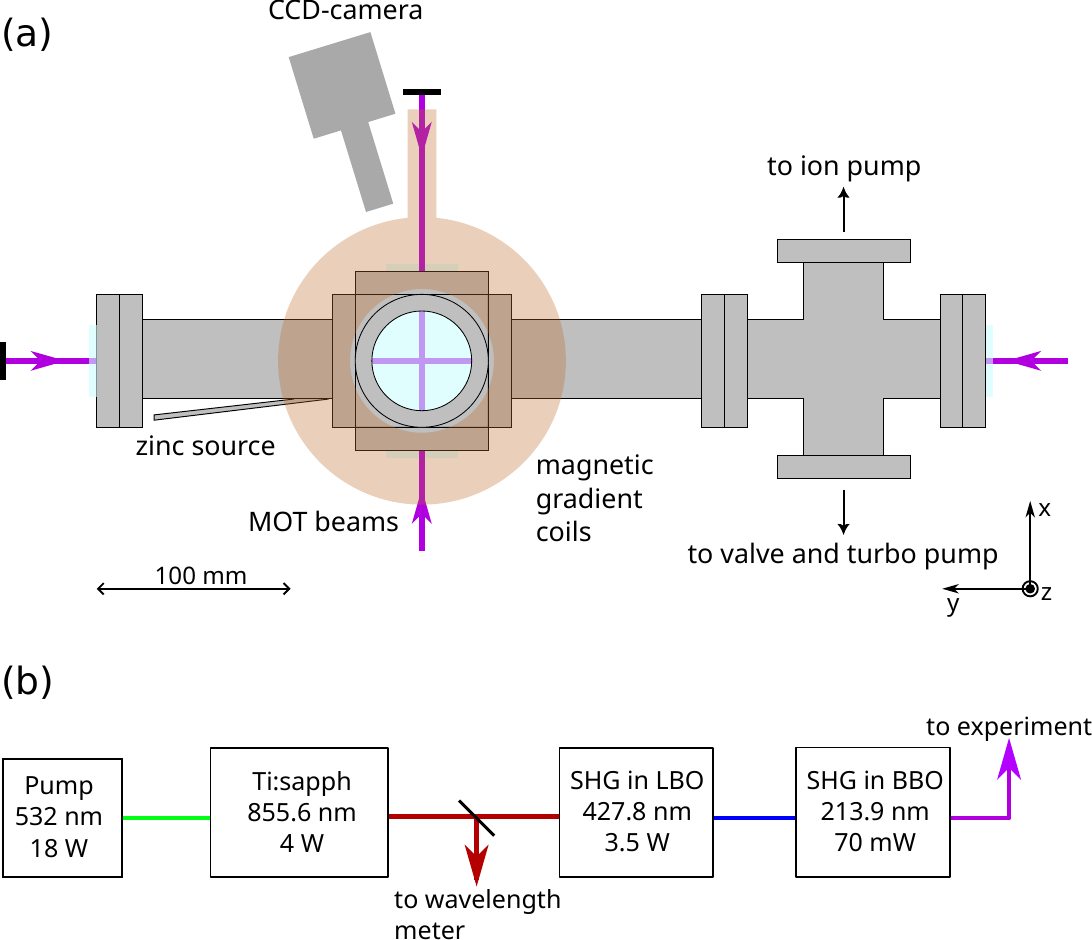}
	\caption{(a) Schematic representation of the experimental chamber. (b) Schematic representation of the 214-nm laser system.}
	\label{fig:setup}
\end{figure}

Fig.~ \ref{fig:transitions} shows a diagram of the relevant transitions. As mentioned above, the DUV transitions of Zn require a powerful and tunable cw laser. Fig. \ref{fig:setup} (b) shows a schematic depiction of the laser system.. We use frequency quadrupling of a titanium-sapphire laser (Ti:sapph) to generate the necessary 213.9 nm light. The Ti:sapph is pumped by a frequency-doubled Nd:YAG with 18\,W output power, generating 4\,W output of NIR light at 855.6\,nm, which is first frequency-doubled in a LiB$_3$O$_5$ (LBO) crystal using an enhancement cavity, with an efficiency of $>80\%$ resulting in up to 3.5\,W at 427.8\,nm. The light is then coupled into a second enhancement cavity containing a BaB$_2$O$_4$ (BBO) crystal. This second cavity uses an elliptical focus at the surface of the BBO crystal to minimize radiation damage while maintaining high conversion efficiencies \cite{Gumm}. One mirror of each cavity is mounted to a piezo, allowing the cavity length to be stabilized to the laser frequency via the Hänsch-Couillaud locking scheme \cite{HC}. This laser system can generate up to 130\,mW of output at 213.9\,nm. During the measurements presented here, the output power was reduced to about $70$\,mW to minimize radiation damage to the BBO crystal.
 
Due to the elliptical focus of the second enhancement cavity, the UV light is strongly divergent. While this divergence is compensated during beam shaping, the resulting beam is still elliptical, with a waist of 6.3\,mm in the major axis and 3.7\,mm in the minor. After beam shaping and collimation the beam is split into three parts, which are retro-reflected after passing the experimental chamber, forming the six MOT beams. Losses on the viewports ($\sim 5\%$ per viewport) lead to a mild power imbalance between counter-propagating beams, which does not impact MOT performance severely. Due to the retro-reflection the light intensity is at the position of the atoms is higher than the input power, which needs to be considered for some of the calculations presented in this paper. For clarity all powers given in this paper are input power, measured before the beam is split into the individual MOT beams.
    
The experimental chamber is depicted in Fig. \ref{fig:setup} (a). The main part of the chamber consists of a CF40 cube. Four of the sides are occupied by viewports, the other two by CF40 tubes. One of these tubes connects to a 4-way cross, allowing for the connection to vacuum pumps as well as another viewport. A small oven containing zinc pellets is welded onto the opposite pipe at an acute angle. This oven is heated by a cartridge heater to control the zinc vapor pressure within the cell. The pressure is monitored and maintained at $4 \cdot 10^{-8}$ mbar by an ion pump.

The large linewidth of the cooling transition ($\Gamma/2\pi = 114$\,MHz) necessitates a large magnetic field gradient of more than 100\,G/cm. A commonly used solution for generating such large gradient fields is to use circular permanent magnets \cite{CdMOT}. However, second-stage cooling on the intercombination line, as well as precision spectroscopy, are incompatible with large and static magnetic fields. We therefore use a pair of large copper coils (40\,mm high, 76\,mm outer radius, 36.5\,mm inner radius, seven windings with 200\,mm$^2$ cross section) to generate the necessary gradient field. The coils are placed around the center cube of the experimental setup in an anti-Helmholtz configuration. They are driven by a high-current power supply at currents up to 2.8\,kA, resulting in a radial gradient field of up to 220\,G/cm. The vertical field gradient is larger by a factor of two. Gradient field strengths given in this paper always refer to the radial field. The coils are attached to water-cooled plates to prevent overheating. The zero point of the gradient field is determined with a Hall probe, and the MOT beams are aligned to this position.

The MOT fluorescence is collected through one of the view ports at the side of the center cube, using a $f=75$\,mm lens in a 4f configuration and recorded using a CCD camera. The recorded photon counts are converted to a corresponding atom number by considering the scattering rate of zinc atoms at the given detuning, optical power and beam size, and calculating the fraction of scattered photons detected on the camera. The number of atoms captured in the MOT is obtained via a two-dimensional Gaussian fit. We estimate that the determination of the MOT population has an accuracy of 50\%, largely limited by poor knowledge of the peak intensities of the MOT beams: while we can determine beam powers with high confidence, the elliptical, non-Gaussian beam shape induces uncertainties in the intensity. Typical MOT populations for $^{64}$Zn range from $10^5$ to $10^6$, depending on the oven temperature $T$, detuning $\Delta$, optical power $P_0$, and magnetic gradient field strength $\mathrm{d}B/\mathrm{d}x = \delta B$. Typical oven temperatures range from 200\,°C to 280\,°C, corresponding to mean atom velocities between 380\,m/s and 430\,m/s.


\begin{figure}[htpb]
	\centering
	\includegraphics[width=0.79\linewidth]{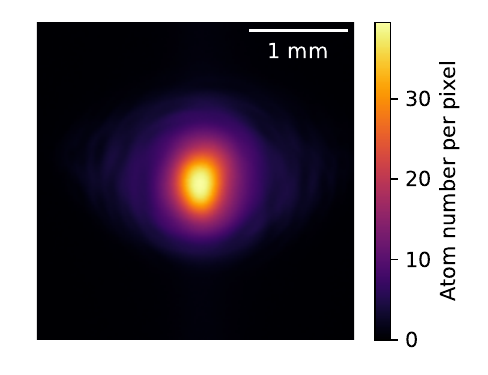}
	\caption{A fluorescence image of a MOT of zinc atoms, as captured by the CCD camera. This image was taken with $P_0 = 62$\,mW, $\delta B = 125$\,G/cm, and $\Delta = 75$\,MHz.}
	\label{fig:example_MOT}
\end{figure}

Fig.~\ref{fig:example_MOT} shows a fluorescence image. The population of the MOT in this image is about $350\,000$. The Doppler temperature of this cooling transition is $T_D=2.7$\,mK, and sub-Doppler cooling mechanisms might be at work for the fermionic isotope, which features a small hyperfine structure in the excited state. Fluorescence images cannot provide reliable information about the MOT temperature.

\begin{figure}[htpb]
	\centering
	\includegraphics[width=\linewidth]{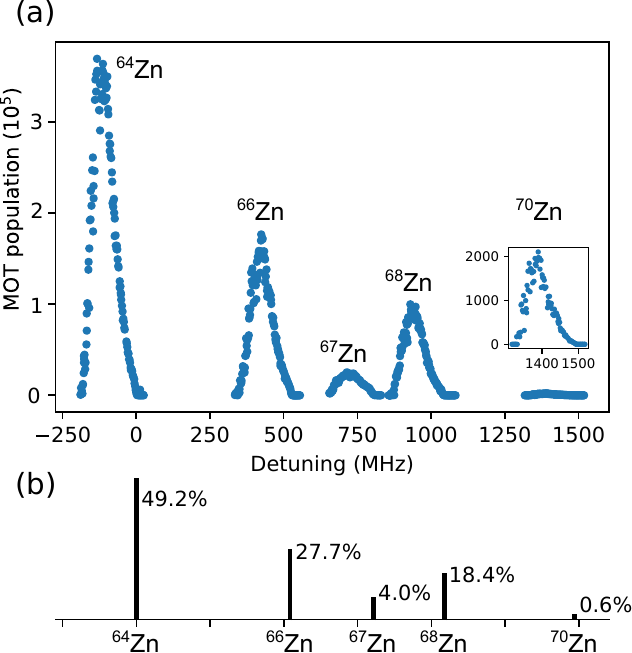}
	\caption{(a) MOT population in dependence of the detuning, given relative to the transition frequency of $^{64}$Zn. (b) Positions of the resonance frequencies for each isotopes, with heights corresponding to the abundance of that isotope.}
	\label{fig:MOT_detuning}
\end{figure}

Fig.~\ref{fig:MOT_detuning} shows the dependence of MOT population on the detuning of the cooling laser. Here, the laser frequency is recorded by a wavelength meter. For the bosonic (even) isotopes, the peak MOT population is reached for a detuning of around $\Delta = -100$\,MHz (or $\sim0.9\,\Gamma/2\pi$).

The fermionic isotope $^{67}$Zn with nuclear spin $I=5/2$ has a $F=5/2$ ground state, and the excited $^1$P$_1$ state splits into three hyperfine states: $F'=3/2$ ($\sim-60$\,MHz), $F'=5/2$ ($\sim-30$\,MHz), and $F'=7/2$ ($\sim50$\,MHz) \cite{roeser}. Here, the values of the hyperfine splitting are given with respect to the center of gravity of the transition. The splittings are similar to the linewidth and to the detuning, leading the light on the $F=5/2 \rightarrow F'=7/2$ cooling transition to be blue detuned to the other hyperfine transitions. As a consequence, the shape of the MOT is noticeably different from the bosonic counterparts, the peak population is found at a detuning of about $-70$\,MHz relative to the $F=5/2$ $\rightarrow$ $F'=7/2$, and the fluorescence peak is less pronounced.
    
\begin{figure}[tpb]
	\centering
	\includegraphics[width=\linewidth]{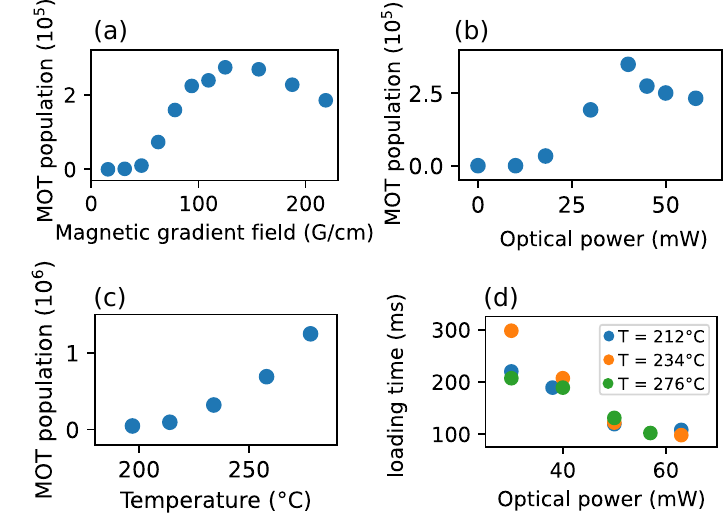}
	\caption{(a) MOT population in dependence of the magnetic gradient field strength, for $T = 230$\,°C, $P_0 = 50$\,mW, and $\Delta = 100$\,MHz. (b) Dependence of the MOT population on the optical power of the cooling laser, for $T = 230$\,°C and $\delta B = 125$\,G/cm. (c) Dependence of MOT population on the oven temperature, for $P_0 = 50$\,mW, $\Delta = 2\pi \cdot 100$\,MHz, and $\delta B = 125$\,G/cm. (d) Loading time in dependence of the optical power and for different temperatures, at $\Delta = 100$\,MHz and $ \delta B = 125$\,G/cm}.
	\label{fig:measurements}
\end{figure}

We measured the dependence of the MOT population on the strength of the gradient field at a detuning of $\Delta = 100$\,MHz, corresponding to the maximum MOT population determined in the previous measurement. The results are shown in Fig. \ref{fig:measurements} (a). The MOT population varies only slightly with the gradient field strength and peaks around $\delta B = 125$\,G/cm. Measurements at different detunings show a similar behavior.
    

Next, we vary the optical power, as shown in Fig.~\ref{fig:measurements} (b). The MOT population increases linearly until reaching a sharp maximum, beyond which the population decreases again. The maximum is reached at $P_0 = 40$\,mW, corresponding to a peak intensity of about 175\,mW/cm$^2$ ($ =0.12\,I_{\rm sat}$) at the position of the MOT, summed over all six beams. The decrease is attributed to photoionization loss.

We measure the MOT loading rate by monitoring the increase in fluorescence shortly after ramping up the gradient field. The time resolution is limited by the refresh rate of the camera. The $1/e$ loading time is obtained by fitting an exponential function to these data sets. Loading times range from $\sim300$\,ms to $\sim80$\,ms, depending primarily on the optical power, and show only little dependence on the oven temperature, as can be seen in Fig. \ref{fig:measurements} (d).
    
We conclude that the lifetime of the atoms in the MOT is mainly limited by photoionization losses and not by collisions with thermal background atoms. Thus we assume that the inverse of the measured loading time corresponds to the photoionization rate $\Gamma_{\rm ion}$, enabling us to calculate the photoionization cross section $\sigma$ via \cite{CdMOT}
\begin{equation}
    \sigma =\frac{\hbar \omega \Gamma_{\rm ion}}{P_e I}.
\end{equation}
Here, $\hbar \omega$ is the photon energy, $I$ is the total intensity of all MOT beams, and $P_e$ is the excited state population in the MOT. We obtain a cross section of $\sigma = 2.0(5)\cdot10^{-22}$\,m$^2$.

In summary, we have presented magneto-optical trapping of zinc atoms on the broad 114-MHz singlet transition at 214 nm, with about $10^6$ atoms at a Doppler temperature of 2.7\,mK. The loading rate and trap lifetime could be improved by addition of a Zeeman slower or a 2D MOT to feed decelerated or pre-cooled atoms into the loading region. Second-stage cooling on the 308\,nm $^1$S$_0 \rightarrow ^3$P$_1$ transition, at a linewidth of about 6\,kHz ($T_r\approx 3 \, \mu{\rm K}$), is the next step towards colder ensembles at higher phase-space density. Such ensembles can then be loaded into a dipole trap for spectroscopy of the clock transition near 310\,nm, where the so-called magic wavelength, at which the polarizability of the ground and excited states are identical, is near 410\,nm.

{Funding - }
We acknowledge funding from Deutsche Forschungsgemeinschaft DFG through grants INST 217/978-1 FUGG and 496941189, as well as through the Cluster of Excellence "ML4Q" (EXC 2004/1 – 390534769), from the European Research Council (ERC) under the European Union’s Horizon 2020 Research and Innovation Programme (Grant Agreement No. 757386 "quMercury"), and from the European Commission through project 101080164 "UVQuanT".

{Acknowledgments - }
We thank D.~Röser and M.~Vöhringer for early experimental work and J.~Domarkas of Eksma Optics for providing dedicated optics. We thank the entire team of the UVQuanT consortium for inspiring discussions and technical advice, especially S.~Truppe from Imperial College London, S.C.~Wright from FHI Berlin, and S.~Hannig of Agile Optic GmbH, for close collaboration.

{Disclosures - }
The authors declare no conflicts of interest.

{Data Availability Statement - }
Data underlying the results presented in this paper are not publicly available at this time but may be obtained from the authors upon reasonable request.


%

\end{document}